\documentclass[manuscript]{acmart}

\usepackage{array}
\usepackage{multirow}
\usepackage{xcolor}
\usepackage{svg}
\usepackage{pdfpages}
\usepackage{threeparttable}
\usepackage{colortbl}
\usepackage{hyperref}
\usepackage{svg}
\usepackage{fontawesome}
\usepackage{tcolorbox}
\usepackage{mdframed}
\usepackage{enumitem}
\usepackage{ragged2e}
\usepackage{caption}
\usepackage{blindtext}
\usepackage{capt-of} 
\usepackage{tabularx} 
\usepackage{float}   
\usepackage{soul}

\definecolor{mygray}{gray}{0.25}
\definecolor{mygray2}{gray}{0.8}
\definecolor{mygray3}{gray}{0.9}
\definecolor{amber}{rgb}{1.0, 0.75, 0.0}
\definecolor{mygreen}{rgb}{0.0, 0.5, 0.0}

\AtBeginDocument{%
  }

\newenvironment{myquote}%
  {\list{}{\leftmargin=0.2in\rightmargin=0.2in}\item[]}%
  {\endlist}

\setcopyright{acmcopyright}
\copyrightyear{2018}
\acmYear{2018}
\acmDOI{XXXXXXX.XXXXXXX}

\acmConference[Conference acronym 'XX]{Make sure to enter the correct
  conference title from your rights confirmation emai}{June 03--05,
  2018}{Woodstock, NY}
\acmPrice{15.00}
\acmISBN{978-1-4503-XXXX-X/18/06}

\begin{document}

\title[Algorithmic Harms in the Child Welfare System]{Algorithmic Harms in Child Welfare: Uncertainties in Practice, Organization, and Street-level Decision-Making}

\author{Devansh Saxena}
\affiliation{%
  \institution{Carnegie Mellon University}
  \streetaddress{5000 Forbes Ave}
  \city{Pittsburgh}
  \state{PA}
  \postcode{15213}
  \country{USA}}
\email{devanshsaxena@cmu.edu}

\author{Shion Guha}
\affiliation{%
  \institution{University of Toronto}
  \streetaddress{140 St. George Street}
  \city{Toronto}
  \state{Ontario}
  \country{Canada}}
\email{shion.guha@utoronto.ca}

\renewcommand{\shortauthors}{Saxena et al.}

\begin{abstract}
Algorithms in public services such as child welfare, criminal justice, and education are increasingly being used to make high-stakes decisions about human lives. Drawing upon findings from a two-year ethnography conducted at a child welfare agency, we highlight how algorithmic systems are embedded within a complex decision-making ecosystem at critical points of the child welfare process. Caseworkers interact with algorithms in their daily lives where they must collect information about families and feed it to algorithms to make critical decisions. We show how the interplay between systemic mechanics and algorithmic decision-making can adversely impact the fairness of the decision-making process itself. We show how functionality issues in algorithmic systems can lead to process-oriented harms where they adversely affect the nature of professional practice, and administration at the agency, and lead to inconsistent and unreliable decisions at the street level. In addition, caseworkers are compelled to undertake additional labor in the form of repair work to restore disrupted administrative processes and decision-making, all while facing organizational pressures and time and resource constraints. Finally, we share the case study of a simple algorithmic tool that centers caseworkers' decision-making within a trauma-informed framework and leads to better outcomes, however, required a significant amount of investments on the agency's part in creating the ecosystem for its proper use.
\end{abstract}




\maketitle

\section{Introduction}
Algorithmic decision-making systems are increasingly being deployed in government agencies such as child welfare \cite{chouldechova2018case, saxena2020child}, criminal justice \cite{stevenson2022algorithmic, stevenson2018assessing}, public education \cite{mcconvey2023human}, unemployment services \cite{charette2018michigan} and homeless services \cite{kuo2023understanding} in an effort to improve public services delivery and target resources toward cases that most need them. Algorithms are being deployed across the public sector where they are informing decisions regarding which child maltreatment cases to investigate \cite{kawakami2022improving, saxena2021framework2}, which families should be offered preventive services \cite{saxena2022unpacking}, which neighborhoods require more policing \cite{brayne2017big, haque2020understanding}, and who are offered public housing and unemployment benefits \cite{eubanks2018automating, kuo2023understanding}, among others. However, audits of these systems show that they are achieving worse outcomes \cite{cheng2022child, robertson2021modeling}, embedding human biases in decision-making \cite{saxena2022train}, and exacerbating racial and socio-economic biases \cite{eubanks2018automating, redden2020datafied, keddell2019algorithmic}. Algorithmic systems are causing real harm to vulnerable people who are witnessing cuts to their unemployment and healthcare benefits \cite{garza_2022, eubanks2018automating}, loss of public housing \cite{Villasenor_2021}, as well as experiencing unfair investigations from the child welfare system \cite{garance_2022, stapleton2022imagining}.

Child welfare (CW) agencies in the United States are increasingly facing severely limited resources, burdensome workloads, and high staff turnover and have turned towards algorithmic systems as they promise to allocate resources more efficiently and fairly \cite{eubanks2018automating, lodato2018institutional, saxena2020conducting} and offer them a means to "do more with less" \cite{redden2020datafied, keddell2015ethics}. These agencies also rely on federal and philanthropic funding to support their operations with more grant funding opportunities being made available to agencies that invest in innovative evidence-based programs and systems reform and integration \cite{simon2023examination}. Recently, more federal funding has been made available to CW agencies that implement the new Comprehensive Child Welfare Information System (CCWIS) data system that further improves data collection mechanisms, integrates data from previously siloed departments, and supports bi-directional data exchange between courts, education systems, and Medicaid \cite{harrison2018tale}. In addition, child welfare agencies have also been the center of public and media scrutiny because of harm caused to children who are removed from the care of their parents \cite{camasso2013decision} but they also receive severe criticism and media attention for child abuse tragedies where the system failed to remove and protect a child \cite{hanson_2021, gajanan_2020}. This has further mounted the pressure on CW agencies to employ algorithmic systems that can assist with high-stakes decisions (e.g., assessing the risk of future child maltreatment events) even though evidence points to such systems disproportionately targeting low-income and minority communities \cite{garance_2022, cheng2022child}.

Critics of AI systems across academia (e.g., \cite{raji2022fallacy, shelby2022identifying}) and the press (e.g., \cite{johnson_2023, Villasenor_2021, garance_2022, garza_2022}) have accumulated significant evidence regarding outcome-oriented harms caused by haphazardly deployed AI systems in the public sector, i.e., disparate harms that are caused to affected communities. Consequently, scholars have begun to synthesize a taxonomy of sociotechnical harms \cite{shelby2022identifying} as well as drawn attention to the fundamental misconceptions about the expected performance of AI systems \cite{raji2022fallacy}. Raji and Kumar et al. \cite{raji2022fallacy} refer to this as the \textit{fallacy of AI functionality} and developed a taxonomy of AI functionality issues \cite{raji2022fallacy}. We build upon this body of work and further draw attention towards \textit{process-oriented harms}, i.e. - harms caused to the fairness of the decision-making process itself as a result of algorithmic decision-making \cite{grgic2018beyond, guerdan2023ground}. In this study, we deepen this discussion about algorithmic harms by examining the impact of algorithmic tools on professional practices, administration at the agency, and street-level decision-making in child welfare. Digital technologies initially altered the workings of public agencies from `street-level' to `screen-level' bureaucracies \cite{bovens2002street}; a transformation referred to as digital era governance \cite{veale2019administration, bullock2018sector}. However, despite promises of significant improvements in efficiencies and cost-effectiveness, several digital tools have fused onto existing street-level discretionary practices without altering public management at a deeper organizational behavior level \cite{veale2019administration, margetts2012information}. A similar exploration needs to take place to assess the impact of algorithmic decision-making on the nature of practice (i.e., street-level human discretionary work), administration (i.e., bureaucratic processes) as well as assess how street-level decision-making is changing and whether algorithms are living up to the promises of cost-effective, consistent, and fair decision-making. In this study, we make the following contributions --

\begin{itemize}[leftmargin=*]
    \vspace{0.1cm}
    \item We show how functionality issues in algorithmic systems can lead to process-oriented harms where they adversely impact the nature of professional practice, and administration at the agency, and lead to inconsistent and unreliable street-level decision-making
    \vspace{0.1cm}
    \item We show how caseworkers are forced to assume the added labor resulting from these algorithmic harms and must conduct \textit{repair work} \cite{jackson2013rethinking} to address the disruption caused to administrative processes and street-level decision-making. This repair work is conducted within the bounds of organizational pressures, time constraints, and limited resources. 
    \vspace{0.1cm}
    \item We highlight the case study of a simple algorithmic tool that centers caseworkers' decision-making processes within a trauma-informed framework and leads to better outcomes for families, however, required a significant amount of investments on the agency's part in creating the ecosystem for its proper use.
\end{itemize}

In the following sections, we first discuss related scholarly work on public sector sociotechnical systems, the current landscape of algorithmic decision-making in the U.S. child welfare system, and the theoretical framework within which we ground our qualitative analysis.


\section{Background}
In this section, we first explain our conceptualization of \textit{algorithms} or \textit{AI} since there is no agreed-upon definition of AI within the AI research community. Next, we discuss related scholarly work regarding digital technologies and algorithmic decision-making in the public sector.


\subsection{What is an "Algorithm"?}
Different definitions of algorithms have been adopted across different disciplines due to the intersection between statistical modeling and machine learning as well as ongoing innovations in advanced neural networks \cite{hill2016algorithm, uliasz2020seeing, seaver2017algorithms}. From a social perspective, we define algorithms through the lens of \textit{street-level algorithms} \cite{alkhatib2019street} - computational predictive tools used to make on-the-ground decisions about human lives and welfare. That is, algorithms that directly affect families utilizing public services and not second or third-removed algorithms that might be internally used by government analytics teams. Technically, we define an algorithm as a computational system that takes in input data, computationally processes data, and produces an output. This output can be a predicted outcome of interest or change in output over time when some statistical property of variables change as is the case with change point detection algorithms \cite{aminikhanghahi2017survey}.  


\subsection{Algorithmic Decision-Making in the Public Sector}
Digital technologies have allowed public services to collect a vast amount of data about citizens in the course of their daily operations in managing and delivering public services \cite{mergel2016big, fernandez2017managing, vigoda2011change, eubanks2018automating}. This has changed governance practices in two distinct ways. First, these systems have improved data sharing practices between different sectors of the government and as Veale and Brass \cite{veale2019administration} note, - these systems have allowed for the digital re-integration of siloed services, data sharing practices aimed at creating a `one-stop-shop' and `end-to-end' service delivery with minimal repeated information gathering. This comprehensive cross-sector data has allowed policymakers, technologists, and academics to narrow their focus on improving high-stakes decision-making by developing data-driven practices that purportedly provide consistent, transparent, objective, and defensible decisions to citizens \cite{saxena2020human, eubanks2018automating, bullock2019artificial}. Algorithmic decision-making in the public sector has generally been adopted in the form of risk assessment algorithms with their primary purpose being the preemptive recognition and mitigation of `risk'; a core organizing concept of this shift in governance and of neoliberal economics \cite{andrejevic2019automating}. That is, improving productivity, accountability, and efficiency by proactively identifying clients in the riskiest circumstances and targeting services towards them using algorithms \cite{redden2020datafied, eubanks2018automating}. This further shift in \textit{digital era governance} that embeds neoliberal principles within public administration has been called \textit{New Public Analytics} \cite{yeung2018algorithmic} and focuses more on risk management based on individual client characteristics while driving attention away from structural and societal problems \cite{redden2020datafied, keddell2015ethics, eubanks2018automating}.

The breadth of research at the intersection of public technology and governance has been wide-ranging, including studies that examine how emergent technologies are shaping collaborative work \cite{bodker2017tying, meng2019collaborative}, designing technologies that empower affected communities \cite{brown2019toward, dombrowski2014government, stapleton2022has, saxena2020collective}, studying issues concerning civic engagement \cite{dow2018between, golsteijn2016sens}, understanding the social and ethical implications of datafied public services \cite{redden2020datafied, holten2020shifting, rask2022data}, as well as uncovering how gig work platforms exploit historical inequities which lead to poor working conditions for data workers and limit their agency. \cite{miceli2022data, posada2022embedded}. Researchers have also unpacked the forms, limits, and complexities of participatory design within the public sector where newer technologies are being designed for the governance of smart cities \cite{whitney2021hci, shadowen2020participatory, heitlinger2019right}. Several of these technologies are being developed through public-private partnerships \cite{dow2018between, lodato2018institutional} where the expectation is that these entities are able to transform data into knowledge and inform decisions that are centered in the efficient allocation of resources \cite{holten2020shifting, rask2022data, nielsen2023cares}. As Holten Møller et al. \cite{holten2020shifting} note, \textit{"here, data becomes the promise of future bureaucratic efficiencies"}. 

Specific to affairs of public administration, researchers have studied how the shifting decision-making latitude has impacted the work of street-level bureaucrats. \footnote{A street-level bureaucrat is a professional service worker (e.g., social worker, police officer, teacher) who operates in the frontline of public service provision. They interact closely with clients and make decisions about them based on how they interpret policies relating to the situations at hand \cite{lipsky2010street}.} Studies have found that value conflicts arise when the logics embedded within the government's digital platforms do not align with street-level bureaucrats' discretion when they tried enacting the same shared values in practice \cite{kawakami2022improving,voida2014shared, dombrowski2014government, holten2020shifting}. Research within public administration, science and technology studies, and human-computer interaction have recently drawn attention to how the digitization of public services is leading to distinct changes in street-level bureaucrats’ discretion \cite{busch2020crafting}, power asymmetries between public officials and citizens \cite{stapleton2022imagining, saxena2022train, saxena2022unpacking}, need for re-skilling of public officials \cite{kawakami2022improving, holten2020shifting, ammitzboll2021street}, and actor transparency in government decision-making \cite{giest2018unraveling}. Lindgren et al. \cite{lindgren2019close} go a step further and argue that "public officials can no longer be understood as merely human". They call for a reinterpretation of citizens' trust in their government in regard to the legitimacy and accountability of e-governments. Busch et al. \cite{busch2018digital} explore public service workers' digitized discretionary practices as they balanced conflicting demands of market-oriented goals and norms of professional practice and found that workers responded positively to digitization when it supported professional aspects of their work. In a similar vein of work, Giest et al. \cite{giest2018unraveling} highlight that a disconnect between bureaucratic processes and digital tools magnified in street-level decision-making where workers' discretionary power was obfuscated and led to more complications and time-consumption in accomplishing daily tasks. In the context of smart cities, Meijer \cite{meijer2018datapolis} further argues that building technologies for governance is centered in political, strategic, and value-laden choices made between three key actors: state, market, and civil society. This requires a re-conceptualization of sociotechnical structures that result from their interactions. Public administration scholars have also foreboded a "digital sclerosis" characterized by the stiffening of governmental processes and lowering of innovation feedback from workers \cite{andersen2020digital}. They predict "decreased bargaining and discretionary power of governmental workers" as one of the early warning signs of this phenomenon. These changes in professional work practices through the adoption of digital tools have similar, yet more serious implications for the adoption of algorithmic tools which further shift discretion away from public service workers; instances of which are already being captured by recent studies in the public sector \cite{kawakami2022improving, saxena2023rethinking}.

As a result, researchers have started investigating the intersection of human discretionary work conducted at the street-level and algorithmic decision-making in public services \cite{alkhatib2019street, holten2020shifting, saxena2021framework2, kawakami2022improving, cheng2022child, ammitzboll2021street}. Alkhatib and Bernstein introduced the theory of \textit{street-level algorithms} to distinctly highlight the gaps in algorithmic decision-making that human discretion needed to address \cite{alkhatib2019street}. Unlike street-level bureaucrats who used discretion to reflexively make decisions about novel cases, street-level algorithms produced illogical decisions that offered no recourse and could only be addressed by `learning’ through new data in the future. Pääkkönen et al. expand upon this theory to highlight that algorithm design needed to identify and cultivate important sources of uncertainty because it was at these locations that human discretion was most needed \cite{paakkonen2020bureaucracy}. Additionally, recent work \cite{ammitzboll2021street, saxena2021framework2, kawakami2022improving} has highlighted the collaborative nature of caseworkers' decision-making processes and the impact of bureaucratic structures that algorithm design needed to account for. In sum, researchers have reached a general consensus that any algorithmic interventions in the public sector needed to understand the complexities of human discretion carried out at the street level when implementing day-to-day bureaucratic processes and legislative policies. Recently, Saxena et al. \cite{saxena2021framework2} synthesized prior work conducted on algorithmic governance systems into a framework for algorithmic decision-making for the public sector which accounts for the complex interdependencies between human discretion, algorithmic decision-making, and bureaucratic processes. Next, we offer a brief overview of this framework and explicate its utility in unpacking the impact of algorithmic decision-making on the nature of professional practice, administration at the agency, and street-level decision-making.


\subsection{Algorithmic Decision-Making Adapted for the Public Sector (ADMAPS) Framework}
Saxena et al. \cite{saxena2021framework2} developed a theoretical framework for algorithmic decision-making in the public sector that accounts for the complex inter-dependencies between \textit{human discretion}, \textit{bureaucratic processes}, and \textit{algorithmic decision-making}. Below, we provide a brief summary and explain the three elements of the framework and their utility for our analysis -  

\vspace{0.1cm}
\textbf{\textit{Human Discretion}} refers to the decision-making process model that practitioners in the public sector engage in where they use their professional expertise, make value judgments, and engage in heuristic decision-making based on the available information. This is especially important in the public sector because government officials must make decisions within the bounds of policies and organizational and resource constraints. Moreover, they must use discretion in resolving missing or conflicting sources of information.

\vspace{0.1cm}
\textbf{\textit{Bureaucratic Processes}} refer to critical governance characteristics that include the systemic constraints within which all decisions must be made, day-to-day protocols, and the legislation that the organization is legally mandated to follow. Bureaucratic processes directly impact practitioners' training and nature of practice (i.e., human discretion) at the agency as well as how well an algorithm is integrated into the day-to-day workflows and decision-making processes.

\vspace{0.1cm}
\textbf{\textit{Algorithmic Decision-Making}} adopts \textit{street-level algorithms} \cite{alkhatib2019street} as a theoretical lens to identify algorithmic systems that are used to make on-the-ground decisions about clients and welfare in the public sector. It draws attention to the relevant data needed for decision-making, the degree of uncertainty associated with predicted decisions, as well as the decision-making latitude (i.e., predictive or prescriptive) that is allocated to algorithmic systems. Authors refer to \textit{algorithmic decision-making} as the most flexible element of the framework that researchers can directly affect by designing systems that balance the other two elements (i.e., human discretion and bureaucratic processes). In addition, the authors further highlight the need to understand human discretionary work because algorithm design needs to make space for (and be preceded by) human discretion to fill in the gaps in data as well as make sense of organizational and legislative protocols.

\vspace{0.1cm}
In sum, the ADMAPS framework showcases how practical trade-offs must be made to manage the cross-dependencies at both the macro- and micro-levels of the algorithmic model to offer autonomy to practitioners and improve human discretionary work. The framework also draws attention to the high degree of uncertainty inherent in the administrative data which consequently means unreliable predictions. Therefore, the goal of algorithms in the public sector must be re-evaluated to support the decision-making processes of stakeholders instead of providing predicted outcomes. In their study, Saxena et al. \cite{saxena2021framework2} focused on the micro-interactions between the dimensions of these three elements to understand why each algorithm failed (or succeeded) to offer utility to child welfare staff and their impact on human discretionary work. However, in this study, instead of focusing on singular algorithmic tools and their impact on human discretion, we draw attention to the broader decision-making ecosystem and critically investigate the macro-interactions between these three elements to assess the impact of algorithmic systems on the nature of practice (i.e.,  interaction between human discretion and algorithmic decision-making), the organization (i.e., interaction between bureaucratic processes and algorithmic decision-making), as well as three-way interactions between the three elements to understand how the nature of street-level decision-making is changing in child welfare.


\subsection{Current Landscape OF Algorithms Used in the U.S. Child Welfare System}
Child welfare (CW) agencies in the United States have increasingly adopted algorithms for high-stakes decisions as they promise to improve decision-making, lower costs, and provide better outcomes to citizens \cite{ringel2018improving, eubanks2018automating}. A nationwide survey on predictive analytics in child welfare conducted by the American Civil Liberties Union (ACLU) in 2021 revealed that 26 states have considered employing predictive analytics in child welfare \cite{samant2021family}. Of these 26 states, 11 are currently using them \cite{samant2021family}. Several states in the United States continue to experiment with predictive analytics, however, audits of these systems reveal that they are achieving worse outcomes for families and exacerbating racial biases \cite{lacounty2017, chicago2017, oregon2022, allegheny2022}. In the past, Los Angeles County and the state of Illinois have shut down their predictive analytics programs for these reasons \cite{lacounty2017, chicago2017} with Oregon recently joining their ranks in June 2022 \cite{oregon2022}. A recent study conducted by Cheng and Stapleton et al. \cite{cheng2022child} on the Allegheny Family Screening Tool (AFST) found that AFST-predicted decisions were racially biased and workers reduced these biases by overriding erroneous decisions. AFST algorithm was designed to mitigate call screeners' biases and subjective decisions and augment decision-making by making it more objective through data. Ironically, AFST has introduced more complexities in decision-making and the call screeners are the ones mitigating algorithmic biases. A comprehensive literature review of algorithms in CWS revealed other sources of biases embedded in the predictors, outcomes, and computational methods being used to develop these systems \cite{saxena2020human}. Kawakami et al. \cite{kawakami2022improving} found that call screeners were offered minimal information about the workings of AFST, considered it to be unreliable due to unexpected model behaviors, and even engaged in a collaborative "game" to learn more about the tool. Gerchick et al. \cite{gerchick2023devil} unpack how design decisions made by AFST's developers effectively act as policy choices that sit outside the purview of democratic oversight. They show how families are labeled "risky by association" because designers aggregated risk scores at the household level and how systemic discrimination is amplified by the inclusion of static variables that capture families' involvement with the criminal legal system or public benefits programs. Similarly, Saxena et al. \cite{saxena2023rethinking} draw attention to the multiplicity and temporality of risk factors that are experienced by families in the system, how the system itself poses significant risk where families experience over-surveillance and are subjected to several arbitrary decisions, and how these street-level structural biases become embedded in algorithmic systems such as the Allegheny Family Screening Tool.

Federal initiatives such as improved data infrastructures for CWS \cite{harrison2018tale} have paved the way for tech startups to develop and pitch algorithmic systems to human services agencies across different states \cite{eckerd, pap, mind-share, sas}. However, there is a need to critically examine the current points of failures in algorithm design as well as understand how workers engage with algorithms as they make critical high-stakes decisions about families. Critical to the conversation about predictive analytics or predictive risk models (PRMs) is also the underlying principle of "risk" and how its understanding has shifted in response to the restructuring of public services to be economically efficient, productive, and accountable \cite{callahan2018paradox, redden2020datafied, andrejevic2019automating}. Traditionally, child welfare services have focused on risks and protective factors within families to be able to provide them with individualized care. However, with a shift towards an economic understanding of risk and the introduction of PRMs, risk has now become a function of client characteristics as existing in prior cases (and not individual family circumstances) and their impact on a predicted outcome (i.e., risk of maltreatment) which Gerrick et al. \cite{gerchick2023devil} refer to families being labeled "risky by association". That is, the risk is estimated based on historical administrative data and is being used to identify the "deserving poor" who pose the most risk to governmental apparatus \cite{eubanks2018automating}. Redden et al. \cite{redden2020datafied} refer to this as the embedded logic of actuarialism that also obfuscates and drives attention away from social and structural issues that bring poor and vulnerable communities under the attention of public services such as child welfare, housing authority, and public assistance \cite{keddell2015ethics}.


\section{Methods}
We partnered with a child welfare agency that serves a metropolitan area in the midwestern United States. This private, non-profit agency is contracted by the state's Department of Children and Families (DCF) to provide child welfare and family services and must comply with all DCF standards, including the use of mandated decision-making algorithms. DCF's Initial Assessment (IA) workers investigate allegations of child maltreatment, and if abuse/neglect is substantiated, the child(ren) may be removed from the care of their parents or an in-home safety plan might be developed. At this point, the case is referred to the agency to provide services. These services are negotiated between the parents' attorney, the district attorney's office, and the judge after caseworkers have conducted initial structured assessments and provided their recommendations to the court. A case manager is assigned to each case and is supported by a multi-disciplinary child welfare team that brings in domain expertise from social work, family psychology, medicine, and law. The agency is mandated to use some algorithms at different stages of a child welfare case per DCF standards but has also developed an algorithm in-house to improve its decision-making processes.


\subsection{Study Overview}
We conducted a two-year extensive ethnography to understand how caseworkers interacted with algorithmic systems, and their perspectives on these systems, as well as unpack how decisions were made at the intersection of child welfare practice, regulations, and algorithmic decision-making. This study is conducted within the same broader child welfare system that was also the ethnographic site for Saxena et al. \cite{saxena2021framework2}. However in this study, instead of focusing on singular algorithmic tools and assessing their impact on human discretion, we draw attention to the broader decision-making ecosystem and assess the impact of algorithms on the nature of practice, administration at the organization, and the changing nature of street-level decision-making. Before conducting observations or recruiting participants for interviews, we obtained Institutional Review Board (IRB) approval at our research institution to conduct our study. Before the interviews, we emailed the participants an IRB-approved consent form and obtained their verbal consent to participate in the study before beginning the interviews. The first author observed 80 child welfare team meetings and then conducted 20 semi-structured interviews with key stakeholders at the agency to better contextualize interactions that occurred at these meetings and gather caseworkers' perspectives. Below, we first highlight our scoping criteria for algorithms followed by a description of observations and interviews, and the qualitative data analysis process.


\subsection{45-Day Staff Meetings, Permanency Consultations, and Concurrent Planning}
45-day staff meetings occur within the first 45 days of a case coming into the care of the agency and are attended by child welfare staff involved at the front end of case management and planning. These meetings facilitate information sharing such that consensus decisions can be made regarding child well-being. Each meeting is scheduled for 90 minutes and the first author observed 25 such meetings. These meetings are typically attended by child welfare staff that work in case management, permanency planning, family preservation, and licensing. The goal of these meetings is to develop a deeper understanding of a family's circumstances from a trauma-informed perspective, identify parents' support system that can help with child care, identify systemic barriers that may inhibit reunification, and develop action steps for child welfare staff. These meetings facilitate collaborative decision-making and ensure congruence in case planning.

Permanency consultation meetings are specialized meetings designed to expedite permanency for children placed in out-of-home care by employing collaborative practices as well as actively addressing any systemic or policy-related barriers. These meetings are facilitated by permanency consultants and are staffed with many of the child welfare team members that attend 45-day planning meetings. These meetings regularly occur at the 5, 10, and 15+ month marks for every case until the case is closed. These ongoing meetings tended to be more informative than the 45-day planning meetings because not enough information is available at the onset of a case. Moreover, permanency consultations involved cases that had been with the agency for several months (if not years) and revealed the complicated interactions between policies, systemic barriers (legal, resource, administrative), social work practice, and algorithmic systems being used at the agency. The first author attended 55 permanency consultations each of which was an hour long.

Critical to a discussion about collaborative meetings and the decision-making ecosystem is the \textit{Adoption and Safe Families Act (1997)} which introduced some of the most sweeping changes to the child welfare system and shifted the focus primarily towards child safety concerns and away from the policy of reuniting children with parents regardless of prior neglect/abuse. The Act introduced federal funding to assist states with foster care, adoption, and guardianship assistance and expanded family preservation services. In addition, it also introduced a \textbf{15-month timeline} where the State must proceed with the termination of parental rights if the child has been in foster care for 15 out of the last 22 months \cite{gossett2017client}. This speedy termination of parental rights has received widespread criticism but still establishes the constraints within which caseworkers must conduct their work \cite{guggenheim2021racial, stewart2022re, polikoff2021strengthened}. To ensure expedited permanency\footnote{Permanency is defined as reunification with birth parents, adoption or legal guardianship.} for foster children, this child welfare agency employs concurrent planning such that two simultaneous plans begin when a child enters foster care: a plan for reunification with the birth parents and a plan for guardianship and/or adoption if reunification is not possible. 


\subsection{Semi-Structured Interviews}
Next, we used the knowledge gathered from these observations to develop our interview protocol and recruit participants who consistently attended these meetings. We conducted these interviews to delve deeper into caseworkers' understanding of these algorithms as well as the benefits and challenges as perceived by them. We asked participants a series of questions about the nature of child welfare work, how algorithms in use impacted their practice, and the organizational support made available to them to mitigate conflicts or challenges associated with these systems. We also asked them to expand upon any interactions we had observed during the meetings such as the participant's dislike or appreciation for a certain algorithm or feature or their frustration with the misuse of algorithmic tools. During observations, we also noted any benefits or frustrations that the participant (or their team) experienced regarding algorithmic decision tools and brought them up during the interview to further expand upon and better understand their outlook. We conducted 20 interviews with participants that included program directors, child welfare supervisors, permanency consultants, caseworkers, data specialists, and clinical therapists.


\subsection{Study Participants}
We interviewed child welfare workers who attended the collaborative meetings mentioned above and had more experience working with algorithmic tools and assessments in their day-to-day work. The case manager position experiences high turnover with most case managers quitting within the first two years. Therefore, we focused our attention on the more experienced members of the child welfare staff who could provide deeper insights into the workings of the system as well as systemic barriers that impede their work. Below, we provide job descriptions of various positions and participant information in Table 1.

Program Directors (n=1): Agency leadership responsible for professional development programs, trainings, research and policy initiatives, and grant writing. They supervise child welfare supervisors and the agency's trauma-responsive service model.  

child welfare Supervisors (n=8): Supervisors manage a case management team comprising 6-8 case managers and oversee about 140 cases. They are responsible for the professional development of case managers and provide additional support when interacting with families and legal parties. They also facilitate 45-day staffing meetings.

child welfare Case Managers (n=2): Frontline workers that directly interact with families and act as mediators between parents, foster parents, relatives, legal parties, and child welfare staff. They conduct home visits, safety assessments, and psychometric assessments, and transport foster children to supervised visits and medical appointments. On average, they manage about 20 cases.

\begin{table}[]
    \small
    \begin{tabular}{|>{\centering}p{1.3cm}|>{\centering}p{0.4cm}|>{\raggedright}p{4.5cm}|>{\centering\arraybackslash}p{1.3cm}|}
         \hline
         
         \textbf{Participant} & \textbf{Sex} & \centering{\textbf{Job Title}} & \textbf{Experience (years)} \\
         \hline
         P1 &  F &  Permanency Consultation Supervisor     & 22     \\ 
         P2 &  F &  Permanency Consultant                  & 20      \\ 
         P3 &  F &  Permanency Consultant                  & 9       \\ 
         P4 &  F &  Permanency Consultant                  & 8       \\ 
         P5 &  F &  Permanency Consultant                  & 12      \\ 
         P6 &  F &  Permanency Consultant                  & 3       \\ 
         P7 &  F &  Child Welfare Program Director         & 20    \\ 
         P8 &  F &  Child Welfare Supervisor               & 19    \\ 
         P9 &  F &  Child Welfare Supervisor               & 13     \\ 
         P10 & F &  Child Welfare Supervisor               & 9      \\ 
         P11 & F &  Child Welfare Supervisor               & 9     \\ 
         P12 & F &  Child Welfare Supervisor               & 7      \\ 
         P13 & M &  Child Welfare Supervisor               & 12     \\ 
         P14 & M &  Child Welfare Supervisor               & 30   \\ 
         P15 & F &  Child Welfare Supervisor               & 9    \\ 
         P16 & F &  Clinical Therapist                     & 5     \\ 
         P17 & F &  Child Welfare Case Manager             & 8      \\ 
         P18 & M &  Child Welfare Case Manager             & 2      \\ 
         P19 & F &  Data Specialist (Program Director)     & 17    \\ 
         P20 & M &  Data Specialist                        & 17     \\ 
         \hline
    \end{tabular}
    \caption{Interview Study Participants}
    \vspace{-0.3cm}
    \label{tab:participants}
\end{table}

Permanency Consultation Supervisor (n=1): The supervisor in charge of the permanency consultation program that is designed to address systemic barriers that pose obstacles in the way of achieving permanency for foster children. They manage and supervise permanency consultants.

Permanency Consultants (n=5): They are responsible for managing the legal process underscoring permanency where they prepare documentation for court, focus on recruitment and licensing of foster homes, and manage post-guardianship and post-adoption services, among other tasks that help expedite permanency for foster children. They provide consultations on about 150 cases each.

Data Specialists (n=2): Responsible for tracking federally-mandated performance benchmarks for the agency as well as case-level data. They also analyze data and manage algorithmic systems implemented at the agency and present results to agency leadership.

Clinical Therapist (n=1): A licensed clinical social worker who conducts mental health assessments and has a deeper understanding of psychometric assessments used in child welfare.


\subsection{Qualitative Data Analysis}
The first author took detailed observational notes during each team meeting and compiled a debriefing document with their initial insights. They also noted questions that needed further clarification and conferred with team leaders (i.e., supervisors or permanency consultants) after the meetings and noted their responses. All authors read through the observational notes and collectively discussed meeting notes that uncovered pertinent aspects of the decision-making ecosystem in regard to the nature of practice, policies and regulations, and/or the use of algorithmic tools. Interviews were audio-recorded and transcribed verbatim for analysis. After carefully reading through the transcripts, we conducted several rounds of iterative coding to identify patterns and converge on appropriate themes associated with the three elements of the ADMAPS framework (i.e., human discretion, bureaucratic processes, and algorithmic decision-making). We performed thematic analysis \cite{clarke2015thematic} to create these initial codes, form a consensus around the codes, as well as resolve any ambiguous codes. In our results, we also use our observational notes to augment the insights we gained from the interviews and note potential discrepancies and nuances from the holistic insights gained from our site observations. These initial codes were grouped into three high-level themes of \textit{impact on practice}, \textit{impact on organization}, and \textit{impact on street-level decision-making}. Codes that highlighted pertinent issues regarding how human discretionary work has changed under algorithmic systems were grouped under \textit{impact on practice}. For instance, 80\% of the participants shared that the CANS algorithm had negatively impacted social work practice due to its lack of understanding of trauma. Similarly, codes that highlighted issues such as an algorithmic system's inability to account for organizational constraints or legislation were grouped under \textit{impact on organization}. For instance, 75\% of the participants shared that all decision-making processes were first situated in the 7ei algorithm which is central to the agency's trauma-responsive service model. Finally, codes that highlighted how street-level decision-making is changing on the ground where bureaucrats are mandated to use algorithmic decisions (but do so while employing their professional expertise and operating under organizational and legislative constraints) were grouped under \textit{impact on street-level decision-making}. For instance, 80\% of participants shared that they were mandated to use the foster care placement as recommended by CANS, however, foster children were more likely to achieve stability and well-being when placed with relatives rather than a more restrictive foster care setting as recommended by CANS. In addition, CANS did not account for the limited number of foster homes available in the system; an organizational constraint that rendered the algorithmic recommendation nonsensical.


\section{Results}
In the following subsections, we first share some overarching findings regarding the decision-making ecosystem in child welfare that sits at the intersection of policies, social work practice, and algorithmic systems. Next, we discuss findings from two algorithms that are in use at the child welfare agency and how their adoption has impacted the nature of practice, administration at the agency, and street-level decision-making.  


\subsection{Decision-making Ecosystem in Wisconsin's Child Welfare System}

\begin{figure}
  \includegraphics[scale=0.26]{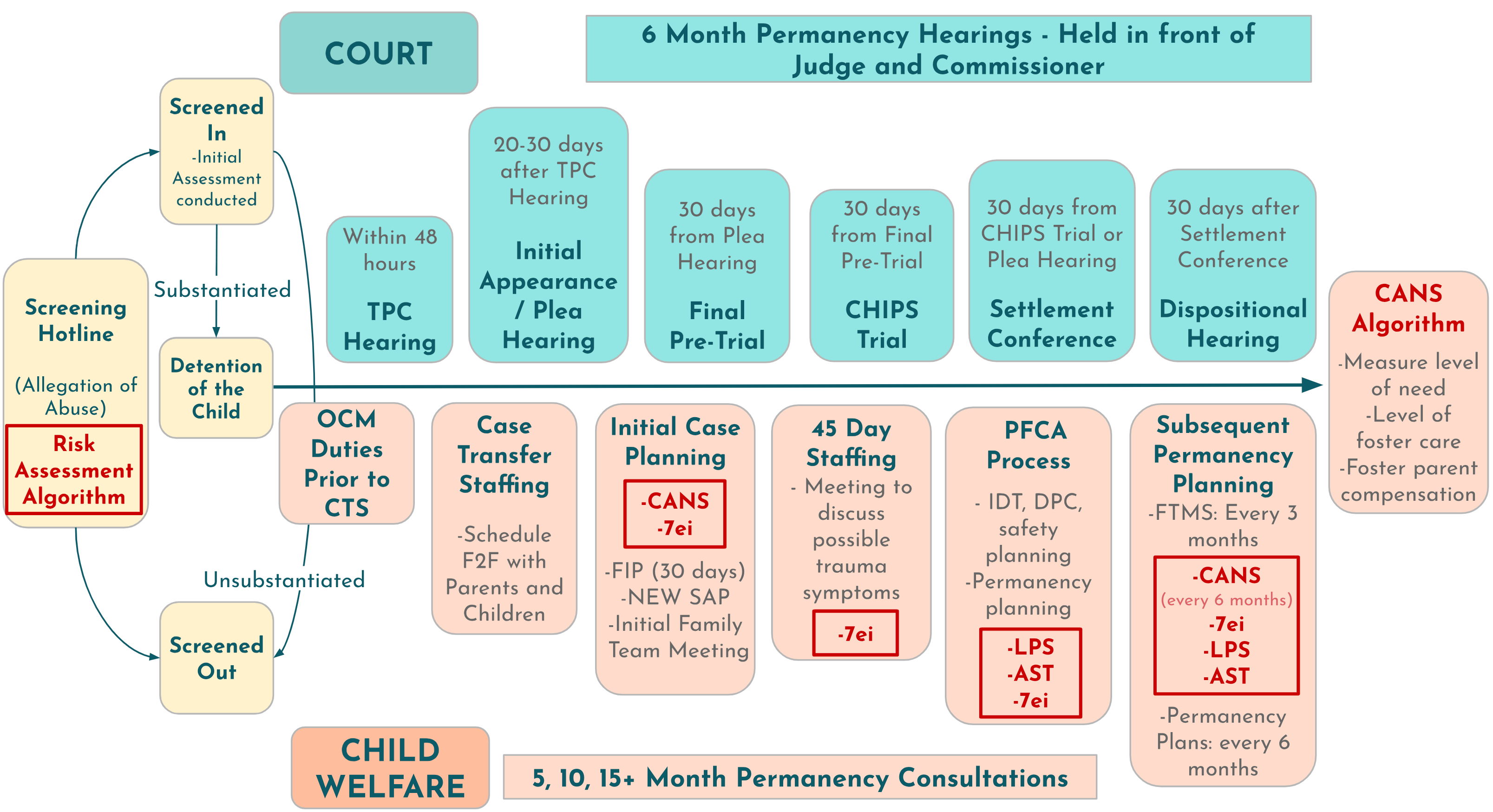}
  \caption{\textbf{Life of a child welfare Case in Wisconsin}}
  \vspace{-0.2cm}
  \caption*{\textcolor{mygray}{IDT: Interdisciplinary team, PFCA: Private Foster Care Agencies, OCM: Ongoing Case Manager, CTS: Caretaker Supplement, TPC: Temporary Physical Custody, F2F: Face to Face, FIP: Family Investment Program, SAP: Substance Abuse Prevention, FTM: Family Team Meetings, DPC: Direct Primary Care}}
  \label{fig:framework}
  \vspace{-0.3cm}
\end{figure}


In order to understand the role of algorithms in decision-making, it is necessary to map out the complexities within the broader decision-making ecosystem to be able to assess the utility and scope of algorithmic tools. In addition, it is imperative to understand the systemic constraints within which such systems must operate. This helps us to better contextualize how caseworkers interact with algorithms, their perspectives on these systems, as well as how decisions are made at the intersection of policies, social work practice, and algorithms, i.e. - how algorithmic decisions were used by child welfare staff working under legislative and organizational pressures. Figure 1 describes \textbf{“Life of a Case”} in child welfare and highlights all the critical decision-making steps and algorithms (red boxes) that are embedded throughout the child welfare process. We co-designed this diagram with caseworkers and it is generally used at the agency to depict the customary legislative process, however, this diagram can significantly vary based on case circumstances. The yellow section of the diagram represents the initial allegation of maltreatment and investigation that is conducted by Initial Assessment (IA) caseworkers at the Department of Children and Families (DCF). If maltreatment is substantiated, the case is officially opened and referred to the child welfare agency to provide ongoing services to the family. This is represented by the orange section of the figure. Finally, the blue section of the figure represents the court system. Contrary to popular belief, critical decision-making power in regard to permanency (i.e., reunification, adoption, or transfer of guardianship) for foster children sits with the court system and child welfare staff only makes recommendations to the district attorney's office. Tensions between child welfare staff and the court system are well captured in social work literature \cite{ellis2010child, duffy2010macro, carnochan2006child}. This ethnography is conducted within the orange section of the figure (i.e., the child welfare agency) where algorithmic tools are being used by caseworkers to make several day-to-day street-level decisions (as opposed to critical permanency decisions made in courts) about children and families. Underscoring the use of algorithmic decision-making in child welfare are two dominant concerns that frequently arose in this ethnography -  

\begin{itemize}[leftmargin=*]
    \item Caseworkers are not trained in “thinking statistically” about data, algorithms, and uncertainties but are legally mandated to input data, interact with algorithms, and make critical decisions.
    \item All algorithmic decisions in the public sector must be made within the bounds of policies, current practice, and organizational constraints.
\end{itemize}

As depicted in Figure 1, when an allegation of abuse is made at the hotline, a risk assessment algorithm helps call screeners to decide if the call should be screened for an investigation. We did not have access to this decision step since it is conducted at DCF before the case is referred to the agency where the ethnography was conducted. If the call is screened in and the investigation substantiates abuse and/or neglect, the child(ren) is removed from the care of their parent(s), or an in-home safety plan is put in place. The case is then referred to this child welfare agency. Throughout the life of the case, the CANS algorithm is used to assess the mental health needs of a child, the level of foster care (level 1 to 5) the child should be placed in, as well as calculate the compensation that foster parents should be paid by the state. 7ei (Seven essential ingredients) algorithm is used to develop trauma-responsive services. LPS (Legal Permanency Status) algorithm is used to track federal benchmarks such as placement stability, as well as policy and systemic barriers. And finally, AST (Anti Sex-Trafficking Response) algorithm is used to assess the risk of sex trafficking for a foster child. These algorithms are being used to make everyday decisions about foster kids but there are also other algorithms that are being used internally by the Department of Children and Families (DCF). For instance, an algorithm is used to assess the risk of re-entry into the system for every case. In this study, we only focus on the CANS and 7ei algorithms.


\subsection{Caseworkers' Perspectives: Assessments or Algorithms?}
Through the observations of child welfare meetings, we learned about several different sources of data that are collected by the agency as a means to provide consistent information about all cases. The agency uses several psychometric assessments for this purpose (see for e.g., \cite{lyons2014measurement, reed2001reliability, lyons1996risk}) as they provide a structured framework to conduct risk assessments. Child welfare initially adopted psychometric risk assessment instruments (RAI) as a means to standardize the process of assessing children's and parents' risks and needs and allow for a more consistent decision-making process. Consequently, RAIs have facilitated the collection of data about children and families for the past three decades. Over the last decade, psychometric and administrative data from prior substantiated cases of maltreatment is now being used to train algorithms to make predictions about current cases \cite{saxena2020human}. The algorithms depicted in Figure 1 have all been developed using their RAI counterparts. RAIs have been widely adopted in social work practice, and as a result, all participants recognized these assessments but did not actively recognize their algorithmic components (i.e., automated aspects) that routinely frustrated them. However, once we focused more on these frustrations and expectation violations, we uncovered several pertinent issues at the intersection of human discretion, bureaucracy, and algorithmic decision-making. In the following sections, we discuss the CANS and the 7ei algorithm and their impact on the nature of social work practice, administration at the agency, and street-level decision-making.


\subsection{CANS Algorithm}
CANS (Child and Adolescent Needs and Strengths) algorithm is constructed using the CANS communimetric assessment that consists of 104 psychometric items organized across eight domains that address child needs and strengths (see \cite{chor2015out} for more details). It was designed to assess the \textit{level of need} of a foster child and utilize this assessment to develop an individualized care plan. CANS offers the child welfare team a structured way to assess a case as well as share this information with other parties. With its primary purpose being communication, CANS is designed based on communication theory rather than psychometric theories centered in measurement development \cite{lyons2009communimetrics}. That is, the tool was not designed to explicitly measure any variables or predict outcomes based on these variables. CANS was designed to be the \textit{expected outcome} of the decision-making process such that it facilitates the linkage between the assessment process and the design of individualized service plans. That is, CANS was designed to support care planning, facilitate quality improvement initiatives, and monitor the outcomes of new evidence-based practices.

CANS algorithm, on the other hand, uses the risks and needs data about children collected by the assessment and has been re-purposed to explicitly measure three outcomes - \textbf{1)} mental health services to be offered to foster children, \textbf{2)} level of foster care they should be placed in, and \textbf{3)} generate subsidized guardianship rate offered to foster parents. Based on a child's risks and needs, the algorithm recommends mental health services that must be offered. Next, based on this level of need, it predicts the level of foster care the child should be placed in. Foster homes in this state range from Level 1 to Level 5. Higher-level foster parents are more trained and certified to take care of higher-needs children. Therefore, the higher the level of need, the higher the level of foster care that the child should be placed in. Finally, based on the level of need and foster home setting, CANS generates the subsidized guardianship rate that foster parents are paid by the state for the costs associated with having a child placed in their care. CANS is conducted within the first 30 days of a child entering the child welfare system or moving to a new placement. Subsequently, it is then conducted every six months. CANS algorithmic assessment is completed by caseworkers with information (about child behaviors and needs) provided by foster parents.

This re-appropriation of CANS (and therefore, CANS data) to predict these outcomes has led to several unintended consequences that frustrate caseworkers and impedes theory-driven practice centered in trauma-informed care. All caseworkers at the agency are trained and certified in conducting the CANS assessment, however, they are not trained in managing conflicts that arise due to the re-appropriation of CANS. This further obfuscates the boundary between the role of the \textbf{assessment} versus the \textbf{algorithm}. Below, we discuss some unintended consequences on the nature of practice, administration at the agency, and street-level decision-making. 

\subsubsection{\normalfont{\textbf{CANS Diminishes the Nature of Social Work Practice.}} \newline} 
In this section, we show how the CANS algorithm adversely affects the nature of social work practice because it does not account for the principles of trauma-informed care, misses important details about the child's mental health, and offers no context regarding worsening behaviors.

\vspace{0.2cm}
\textit{\textbf{CANS contradicts trauma-informed care.}} Caseworkers at the agency are trained in trauma-informed care (TIC) and apply principles and practices from TIC while employing the agency's trauma-responsive service model to cases. However, the CANS algorithm only offers a live snapshot of the child's mental health based on exhibited behaviors (over the past 30 days) and does not account for underlying trauma (or traumatic triggers) that can cause serious emotional dysregulation from time to time. Here, caseworkers co-opt CANS to account for trauma-informed care and anticipate the child's needs. One permanency consultant explained - 
\begin{myquote}
 \small{\textcolor{mygray}{"We have a child that goes into manic depression every year around the holiday season. At this point, one of the foster parents has to quit their job and care for the child full time. We know that this traumatic trigger is coming up but there is no way to account for that in CANS. So we edit scores in anticipation for these upcoming needs"} -- P3, Permanency Consultant}
\end{myquote}

\textit{\textbf{CANS misses important context about the child's mental health.}} Algorithms attempt to generalize child and family characteristics and place them in certain categories to be able to make a determination. However, this inadvertently leads to a loss of information or context since all the information cannot be accounted for. CANS focuses on child behaviors, risks, and needs but it does not account for the child's interactions and relationships with other people. Caseworkers shared that CANS conducted the child's assessment in an isolated manner and did not account for the quality and impact of relationships in their lives which are often more important in determining their long-term well-being. For instance, a permanency consultant shared - 
\begin{quote}
    \small{\textcolor{mygray}{"A child's behaviors are often a result of what is happening in their environment. What or who is really triggering them and where is that coming from.. there is no way to put that in CANS"} -- P4, Permanency Consultant}
\end{quote}

\textit{\textbf{CANS offers no context regarding worsening behaviors.}} A clinical therapist shared that it was more important for them to be able to understand the context around worsening behaviors. They shared that worsening behaviors were not always a bad thing since children's behaviors initially get worse (before getting better) when they started therapy. This is simply because children are finally starting to address the underlying trauma in their lives in therapy which might lead to worsened behaviors in the short-term but is necessary for their mental health and well-being in the long run. The clinical therapist also shared the following concern about the short-term focus on mental health - 

\begin{quote}
    \small{\textcolor{mygray}{"If CANS is used as a communication tool, that is, just as a structured way to talk about children and families and then to make decisions on it, then that's fine! But that's not how it's being used.. A lot of it is just entering [data] into a computer and seeing whether or not there's improvement from one point to another. And if there's no improvement then making a decision." }-- P16, Clinical Therapist}
\end{quote}

These concerns regarding the re-appropriation of CANS were inconsistently shared by child welfare workers. That is, not all caseworkers were equally aware of other concerns and only shared issues that affected their day-to-day practice. For instance, caseworkers were unaware of the clinical therapist's concerns regarding \textit{worsening behaviors}. A supervisor shared that children can have severe underlying trauma that is not captured by CANS because of its focus on exhibited behaviors. At this point, we shared the clinical therapist's concerns with the supervisor and they further elaborated by saying that underlying trauma did not always manifest in terms of exhibit behaviors and that CANS may be measuring the wrong indicators altogether. This also raises questions regarding which explanations associated with the tool are deemed more useful - caseworkers were more invested in explanations of underlying trauma (based on their training in trauma-informed care), whereas the clinical therapist wanted better explanations about the context surrounding worsening behaviors. That is, different stakeholders, have different needs in regard to explanability and a “one-size-fits-all” approach with respect to algorithm design may not be feasible here. 

Such interactions with the algorithm over time led to cumulative distrust in algorithmic decision-making. The implementation of CANS also further obfuscated the difference between the algorithm and the assessment. For instance, caseworkers generally score CANS on the paper-based assessment when speaking with foster parents. However, at the agency, they input this data into the CANS algorithm to predict outcomes. Re-appropriation of CANS is also a case in point regarding why it is imperative to understand the context within which (and how) data is collected about clients in the public sector.

In sum, CANS diminishes the nature of social work practice centered in trauma-informed care because it misses important context about children's mental health, does not account for the interactions in their ecosystem, and offers no context regarding worsening behaviors. Caseworkers' ongoing interactions with the algorithm where these conflicts continually arise have led to a cumulative distrust in the algorithm. 

\subsubsection{\normalfont{\textbf{CANS has Introduced More Barriers to Administrative Processes at the Agency.}} \newline} 
Below we discuss the impact of the CANS algorithm and its re-appropriation on the day-to-day bureaucratic processes carried out at the agency. CANS has added more barriers to day-to-day administration where caseworkers are forced to provide added labor to redress these barriers.

\vspace{0.2cm}
\textit{\textbf{Subsidized guardianship rate has become the primary outcome of interest.}} Both caseworkers and foster parents are actively aware that the subsidized guardianship rate is directly tied to mental health needs. This leads to the gamification of the algorithm for three main reasons - 

\begin{itemize}[leftmargin=*]
    \item Maintaining a stable placement - There is a lack of good foster homes in the system and caseworkers try to support foster parents by any means necessary to ensure that the placement is not disrupted and the child does not need to be placed elsewhere. Multiple placement moves for foster children are associated with poor well-being outcomes where they are unable to develop any meaningful relationships with foster parents or other caregivers \cite{blakey2012review}. Therefore, gaming the algorithm is the only way in which placements can be continually supported.
    \item Base rate is low - Caseworkers shared that the subsidized rate for most foster placements was too low and there was no financial incentive for foster parents to be doing this work. Here, gaming the algorithm is a convenient way to generate a higher compensation by exaggerating the mental health needs on the assessment.
    \item Maximizing financial incentive - Caseworkers also shared (see quote below) that even though most foster parents did their best to care for foster children, there were still other foster parents who accepted multiple placements in their home and ran foster care like a business. They routinely exaggerated child behaviors to receive higher compensations and renegotiated to what they considered to be their "standard" rate. One supervisor explained -   
\end{itemize}

\begin{quote}
    \small{\textcolor{mygray}{"Case managers are being pressured into scoring children higher. Foster parents will match [the rate] with the previous kids. They think that their previous kid got 1500 dollars, so now that is their standard rate, and will demand 1500 dollars for the next kid. They don't have an understanding of CANS or the child's strengths or needs. Money is the key part of these decisions."} --P13, Child Welfare Supervisor}
\end{quote}

We refer to caseworkers manipulating the CANS data as \textit{gamification} to draw attention to the fact that inexperienced caseworkers socially learn from seasoned caseworkers how to play with CANS data in order to generate higher compensations without raising any red flags. However, \textit{repair work} \cite{jackson2013rethinking} provides a better conceptual lens for capturing the shift in the nature of practice because caseworkers must provide the added labor to redress the disruption in decision-making caused by the CANS algorithm. Moreover, the intent behind conducting this repair work is to make the algorithm work for the clients and not just meet the allocation demands of policymakers.

\subsubsection{\normalfont{\textbf{CANS has Added Inconsistencies to Street-level Decision-Making.}} \newline} 
In this section, we discuss CANS' impact on street-level decision-making. Instead of making decision-making more evidence-based, CANS has introduced arbitrariness to the process where 
children are being sent to unnecessary services and the data no longer reflects the true mental health needs of children. We discuss the impact of the CANS algorithm on street-level decisions that caseworkers make in their work relationships with foster parents.

\vspace{0.2cm}
\textit{\textbf{CANS punishes good foster parents.}} Even if there were safeguards in place that prevented the gamification of CANS, the core incentive structure of the algorithm is problematic. Foster parents offer stability and support so that kids can develop good coping skills. They also help address mental health needs and help kids stabilize by taking them to therapy and all their activities. However, CANS scores are recalculated every six months and as the child supposedly exhibits less overt behaviors, their needs are lowered (per CANS assessment) and so is the compensation offered to foster parents. One supervisor explained (see quote below) that by lowering the rate, foster parents were being punished for being actively involved and caring for foster children. The CANS algorithm is set up to incentivize worsening behaviors.

\begin{quote}
    \small{\textcolor{mygray}{"There are an enormous amount of foster parents that do an exceptional job. And so, it is hard, how do you reward them? Because in my opinion, it needs to be a rewarding system... like you're putting in the energy.. you're doing what you're supposed to do... There are foster parents getting their kid into all these activities and making sure this kid has a "normal social life", like normalcy, just part of parenting law. And they are getting half the rate compared to foster parents who are putting in no energy."} -- P9, Child Welfare Supervisor}
\end{quote}

\vspace{0.2cm}
\textit{\textbf{Inconsistent placement decisions due to a lack of foster homes.}} All participants shared that the majority of placement decisions (i.e., where to house a foster child) come down to the availability of good foster homes. For instance, a child might have severe mental health or medical needs and the algorithm might recommend placing the child in a residential care center. However, most of these centers have limited openings, and the child welfare staff is forced to manipulate CANS scores and place the child in a group home or foster care setting that is ill-equipped to manage their needs. In addition, policy dictates placement decisions regarding relatives willing to accept guardianship for a child. For instance, a supervisor shared that relatives' homes must meet all legal and safety requirements and go through a licensing process to become foster parents (see quote below). However, if relatives fail to meet these requirements, CANS is used again to recommend a placement setting where the foster child can at least be temporarily placed until better placements become available.

\begin{quote}
    \small{\textcolor{mygray}{Houses need to be legally compliant with all safety codes before a child can be placed there. These rules make sense on paper but in practice, we lose so many good placement options because relatives can’t afford to move or fix everything in the house to be legally compliant" } -- P10, Child Welfare Supervisor}
\end{quote}

In sum, the re-appropriation of CANS has become a source of frustration for child welfare staff due to inconsistent street-level decisions where they must continually find ways to address policy and systemic barriers as well as ways to make the algorithm work for their clients.


\subsection{7ei (Seven Essential Ingredients) Algorithm}
This child welfare agency's core service delivery model is centered in trauma-informed care (TIC) where the staff assesses every case from a trauma-informed perspective and develops trauma-responsive interventions. However, program directors at the agency were concerned that decisions were still being made in an arbitrary manner where only some child welfare teams were employing TIC practices. To address these gaps in practice, agency leadership has made significant investments towards developing a comprehensive four-part evidence-based program. The program is based on core concepts of the Neurosequential Model of Therapeutics \cite{perry2009examining}. Recently, principles of trauma-informed care have also been adopted within computational research where Chen et al. \cite{chen2022trauma} have proposed a framework for trauma-informed computing that translates these principles into practical design guidelines. 

The 7ei algorithm (commonly referred to as the 7ei tool) is one of the four parts of this program. It is designed using TIC principles where the child welfare team discusses and scores variables across seven domains. We provide some high-level explanations of the seven domains in Table 1 but do not have the agency leadership's permission to share the complete tool. Instead of predicting a singular outcome of interest, 7ei is used to assess the trajectory of child welfare cases, i.e. - \textit{change in 7ei domains over time}. Supervisors and program directors use 7ei to monitor trends in the seven domains to ensure progress is being made in cases using a seven-pronged TIC approach. For instance, if supervisors notice a downward trend for \textit{Regulation} for a given case, it triggers consultations with a clinical supervisor and the development of consequent action steps that must be followed. This integrated approach allows the team to also focus on the child's and parent's ecosystem and their social support system such that parents have more caregiver support in the future. In addition, casenotes from the child welfare team are also collected and linked to the 7ei quantitative scores which provide program directors with more contextual information about the scoring of variables at any point in time. This has allowed the agency to collect pertinent data about their practices in TIC and informs future improvements towards their TIC service delivery model. Below, we discuss how 7ei has impacted the nature of social work practice, administration at the agency, and street-level decision-making.

\begin{table}[]
\Small
\begin{tabular}{>{\raggedright}p{2.5cm}|>{\raggedright\arraybackslash}p{10.5cm}}
\hline
\textbf{7ei Domain} & \textbf{Explanation}\\
\hline
Prevalence & Exposure to and difficulty adjusting to adverse life experiences, i.e. - what happened to the child regarding trauma exposure?\\
Impact & Trauma occurs when a person’s ability to cope with an adverse event is overwhelmed and contributes to difficulties in functioning\\
Perspective Shift & Caregivers' understanding of child's trauma exposure and its relationship with child's behaviors and/or impairments\\
Regulation & Traumatic triggers that cause serious dysregulation and action steps to address it\\
Relationships & Current progress towards facilitating strong relationships in a child's life. Strong relationships help create resilience and mitigate effects of trauma\\
Reason to Be & Child's sense of identity and purpose and their sense to connectedness to their family, community, and culture \\
Caregiver Capacity & Caregivers' understanding of their importance to this process and whether they have additional social support\\
\hline
\end{tabular}
\caption{7ei Algorithm: Explanation of Seven Domains.}
\vspace{-0.5cm}
\label{tab:casenote_example}
\end{table}

\subsubsection{\normalfont{\textbf{7ei Augments the Nature of Social Work Practice.}} \newline} 
\vspace{0.1cm}
The first and second components of the four-part program involve extensive ongoing TIC trainings where the child welfare staff (e.g., caseworkers, supervisors, family preservation team, permanency consultants) is introduced to the complexity of trauma, frameworks for understanding the effects of trauma, and practices and principles of TIC. This is accompanied by specialized supervision and consultation with a clinical supervisor, caregiver support specialist, program administrator, and a national expert (Dr. Bruce Perry) who provide varying degrees of support to frontline caseworkers based on family circumstances. These two components ensure that the nature of practice is centered in the TIC model and allows for deeper engagement with 7ei. One supervisor explained - 

\begin{quote}
    \small{\textcolor{mygray}{"Every case is centered in the trauma-responsive model using 7ei. It ensures that caseworkers are always thinking through TIC"} -- P14, Child Welfare Supervisor}
\end{quote}

Even though most participants shared that they appreciated 7ei and the trauma-responsive model, some participants drew attention to deeper systemic issues in child welfare and the nature of social work practice that need redressing for tools like 7ei to offer more utility. One program director explained - 

\begin{quote}
    \small{\textcolor{mygray}{"During home visits, caseworkers need to ask the right questions and read situations to be able to derive meaningful information. But we are not hiring people who have been doing this for 5-10 years or are highly qualified. The only new hires available are newly graduated social workers." } --P7, Program Director}
\end{quote}

In sum, 7ei was not designed to address such issues in child welfare, however, it is pertinent to note that any algorithmic tool can only offer limited utility when the workers' decision-making latitude (and social work practice) are inherently impacted by deeper systemic and structural issues. That is, the tool augments the quality of social work practice itself but it must still operate within the restrictive legislative framework of the child welfare system \cite{saxena2023rethinking}.

\subsubsection{\normalfont{\textbf{7ei Required the Development of New Administrative Processes at the Agency.}} \newline} 
\vspace{0.1cm}
The third component introduces a staffing protocol in the form of specialized collaborative meetings (commonly referred to as \textit{7ei meetings}) where experienced members of child welfare staff (e.g., family preservation, program director, permanency consultants) share their expertise and provide support to frontline staff members (i.e., caseworkers and supervisors). One supervisor explained - 

\begin{quote}
    \small{\textcolor{mygray}{"I get the most out of having those conversations. What usually happens is that we end up talking about other things, but then it [7ei] brings us back around then too. So, we may be talking about \textit{Prevalence} and \textit{Impact} but when we are on \textit{Relationships} or \textit{Regulation}, we can start tying those to \textit{Prevalence} and how we may be able to help with an intervention or explain why \textit{Regulation} is off. So, I like connecting the dots through TIC and having conversations and processing it [7ei] with my staff."} -- P8, Child Welfare Supervisor}
\end{quote}

Implementation of 7ei (i.e., the fourth component) has also helped the agency address some problems in CWS. Lack of supervisory support is one of the main reasons for high turnover in child welfare \cite{carnochan2013achieving}. 7ei meetings ensure that new caseworkers are receiving adequate supervision and support from experienced members where they are also consistently using the trauma-informed framework. Agency also decided to train only one member of the child welfare team (permanency consultant or supervisor) on the definitions and scoring criteria for variables where they facilitate the meetings and utilization of the tool. This ensures that the rest of the team is able to freely brainstorm as they work through TIC principles. Moreover, any decision-support tool runs the risk of becoming a safe default in a high-stakes environment where the workers carry high workloads and shifts decision-making accountability away from the workers. Here, the 7ei meetings provide the adequate time and space for the tool's collaborative and ethical use.  

\vspace{-0.1cm}
\subsubsection{\normalfont{\textbf{7ei Improves Street-level Decision-Making within Systemic Constraints.}} \newline} 
\vspace{0.1cm}
7ei algorithm is constructed using TIC principles and is embedded within the agency's service delivery model that child welfare staff must follow. One supervisor shared that the caseworker position experienced high turnover such that the agency was always understaffed. Here, it becomes imperative that the agency is creating mechanisms to train new caseworkers where they are employing evidence-based best practices in street-level decision-making when they work with families. 7ei algorithm and 7ei meetings have helped create this training process - 

\begin{quote}
    \small{\textcolor{mygray}{"7ei is embedded in the agency. Everything is based in the trauma-responsive model and often caseworkers are working through 7ei domains without even realizing it"} -- P9, Child Welfare Supervisor}
\end{quote}

However, participants also shared that even though 7ei is leading to better outcomes for some cases, it also ends up being impractical for cases where critical decisions come down to systemic or legal barriers. For instance, a family may have completed trauma-informed interventions and made changes to their household to provide a safe environment for children, however, the district attorney's office might still advocate against reunification because of the family's past (e.g., criminal history, domestic violence, drug use). One supervisor explained - 

\begin{quote}
    \small{\textcolor{mygray}{"7ei is useful but so many decisions come down to systemic problems. Like..for example..the biggest barriers can come from legal parties. They are not in the [parent's] home every month. They are not talking to these parents, day in and day out. So, sometimes it can be challenging to try and fight for reunification."} -- P12, Child Welfare Supervisor}
\end{quote}

Here, the supervisor alludes to ongoing systemic tensions between child welfare staff and the court system that are well-documented in social work literature \cite{ellis2010child, duffy2010macro, carnochan2006child}. 7ei was not designed to address these systemic issues, however, it still adds to the frustrations of caseworkers who continually employ trauma-informed practices using 7ei but are unable to receive a favorable decision in court for their clients.


\section{Discussion}
In this section, we discuss how algorithmic tools at the agency are inadvertently causing harm to the nature of practice, harm to the administration at the agency, as well as uncertainties in street-level decision-making. Next, we discuss implications for human-AI decision-making and implications for responsible data science practices in the public sector.


\subsection{Algorithms Harms to Social Work Practice}
The child welfare system has traditionally suffered from inconsistent decision-making with respect to child safety and family maintenance \cite{camasso2013decision}. The problem of inconsistent decision-making is further aggravated by the fact that CWS experiences chronic turnover with the majority of caseworkers quitting within the first two years \cite{shim2010factors, barak2006they}. However, research in evidence-based social work suggests that it takes caseworkers about two years to learn how to do the job and adeptly navigate and interact with the legal parties, service providers, children, and families, and learn the nature of child welfare practice \cite{edwards2018characteristics}. Consequently, inexperienced caseworkers who use risk assessment algorithms such as CANS are led to believe that they are acting in an unbiased and objective manner. But as our results in this study and prior work \cite{saxena2022train} indicate, bias can be embedded within these assessments and the underlying data collected as a result of the practices of similarly inexperienced caseworkers. For instance, two of the most significant predictors of risk of maltreatment per the WARM risk assessment are \textit{parents' cooperation with the agency} and \textit{stress of caretaker} \cite{saxena2022train}. These variables are singly scored by caseworkers and encapsulate their impressions of the family with no input from the parents themselves. In addition, such variables capture a parent's response to the agency's intervention in their lives rather than the effectiveness or means of the intervention itself (i.e., how the caseworker approached and engaged with the family). child welfare is supposed to transition towards a "families as partners" model, however, algorithmic tools are creating a layer of obfuscation where such power asymmetries can become embedded within these tools. Ironically, risk assessments have worsened the turnover problem in CWS. Caseworkers leave due to their frustrations with practices and working conditions that have been exacerbated by the adversarial nature of risk assessments \cite{callahan2018paradox}. In addition, what remains missing from the conversations about algorithms in child welfare, and the public sector in general, is the significant amount of data labor that is required of caseworkers as well as the repair work \cite{jackson2013rethinking, watkins_art, kellogg2020algorithms} necessary to bring these systems to work for families and not just the governmental apparatus. 

In addition, caseworkers at the agency were frustrated that the CANS algorithm required them to provide data labor where they must collect information about children and feed it to the system, however, the system also stripped them of discretionary power in regard to these decisions. As highlighted by recent work \cite{hong2022prediction}, predictive systems in high-stakes domains are \textit{extractive by design} where they lead to a systemic extraction of discretionary power such that probabilistic outcomes are being used to supplant workers’ contextual knowledge. Moreover, this interaction between caseworkers and algorithmic tools is further problematic because \textit{human-in-the-loop} is the often proposed solution to erroneous decisions made by algorithms \cite{de2020case}. However, in this scenario, humans may be just as likely to make mistakes. Power asymmetries and incentive structures also directly impact how data is collected about children and how these assessments are scored. For instance, the CANS algorithm is conducted by both clinical therapists and caseworkers. However, a prior study \cite{lyons2004measurement} conducted on CANS found that clinical therapists were more likely to detect mental health needs because they are medically trained to detect these needs but also because they provide services when needs are detected, and consequently, they are paid for providing these services. That is, there was a clear financial incentive to exaggerate CANS scores. On the contrary, caseworkers very significantly less likely to detect mental health needs because `detecting needs' inadvertently created more work for them. Once needs are detected, caseworkers must reach out to service providers and secure appointments for their clients.


\subsection{Algorithmic Harms to the Child Welfare Agency}
child welfare agencies in several states in the United States have continued to rely on counsel from the federal government in the form of initiatives, regulations, and evidence-based approaches to improve their practice model. As previously noted, recent federal initiatives have laid the groundwork for more algorithmic interventions in CWS. However, federal directives have continually focused on the need for CWS agencies to adopt data-driven practices without providing adequate guidelines that focus on the \textbf{\textit{why}} and \textbf{\textit{how}} CWS agencies could employ evidence-based data-driven practices within their day-to-day processes in addition to training caseworkers on these practices. Consequently, CWS agencies in several states have rushed to adopt “something” in order to prove that they are employing scientific and evidence-based practices without ensuring that child welfare stakeholders have a strong understanding of how the model works, how to assure fidelity, and how to assess the model for issues of ethics and equity. The Allegheny Family Screening Tool (AFST) is a case in point of this scenario. A recent study conducted by Kawakami et al. \cite{kawakami2022improving} found that call screeners were offered minimal information about the working of AFST, considered it to be unreliable due to unexpected model behaviors, and even engaged in a collaborative "game" to learn more about the tool. Moreover, Cheng and Stapleton et al. \cite{cheng2022child} found that AFST-only decisions were racially biased and workers mitigated these biases by overriding erroneous decisions. Yet, these (sometimes hastily adopted) data-driven risk assessment models have become a central activity in many child welfare organizations \cite{callahan2018paradox}. As child welfare agencies across the United States begin to adopt the new CCWIS data model \cite{harrison2018tale}, it has also paved the way for tech startups \cite{eckerd, pap, mind-share, sas} to start pitching CCWIS-based algorithmic tools to agencies to help them meet their federal and state accountability requirements \cite{corrigan_2023}.

At this agency, there are serious data provenance concerns about data collected about children through the CANS algorithm since the data is so heavily manipulated by both caseworkers and foster parents. CANS was reappropriated to calculate foster parent compensations because policymakers believed that it would offer a fair and unbiased means for allocating resources. In addition, it would reduce costs over time since the improvement in care would lower the mental health needs and also the resources required to provide care. However, data specialists at the agency shared with us several cases where the compensation continually increased over time because the algorithm was being gamed. Another unintended consequence of exaggerating CANS scores is that foster children are now being sent to services (e.g., individual therapy) that they don't necessarily need. This is an added financial burden on an underfunded system that must pay for these unnecessary services. Here, CANS has added more barriers to consistent and evidence-based decision-making and introduced more constraints within the process. Moreover, caseworkers are mandated to provide the data labor that runs AI systems, however, they have no agency over these data production or future data utilization processes. This non-profit agency is contracted by the state's Department of Children and Families to provide child welfare services and the agency must use DCF's centralized data system (i.e., eWiSACWIS) to record all case data. However, they are unable to download this data from their cases for the purpose of critiquing and improving their own practices. Skeptics of AI have invoked ethical frames where engineers and designers of algorithms must refuse to design systems that may raise social and ethical concerns \cite{barocas2020not}. However, such concerns might not arise at the onset, and as Selbst et al. note - "repurposing algorithmic solutions designed for one social context may do harm when applied to a different context" \cite{selbst2019fairness}. CANS algorithm is a case in point of this scenario which was developed by academics with all the best intentions and as a medium to facilitate sharing of information about children's needs. Refusal to design may also not be an option for most public sector agencies who must continually prove that they are employing innovative data-driven practices and may also be under political pressure to adopt algorithmic tools that seek to automate public service delivery as well as measure performance standards that impact future contracts. 


\subsection{Algorithmic Harms to Street-Level Decision-Making}

\begin{quote}
    \small{\textcolor{mygray}{“All government policy and regulation is contradictory. That’s why we exist, to fill in these gaps and make sense of policy in implementing it. This CANS is basically policy, right? So we interpret it and bend it however necessary to make it work for the people. I don’t think we are gaming the algorithm. Because that would mean we are gaming all policy.”} -- P7, Program Director}
\end{quote}

With inadequate federal guidelines on how to adopt, implement, and use algorithmic tools, there is significant misunderstanding regarding the roles of these tools in implementing rules or standards. Rules can be thought of as triggering criteria, i.e. - authoritative conditions that can be algorithmically coded for. On the other hand, street-level bureaucrats have traditionally exercised a significant amount of autonomy and flexibility in applying professional standards. CANS algorithm offers a case in point where it is expected to be implemented as a rule, however, caseworkers engage with it as if it were a standard. In addition, the program director in the above quote alludes to the \textit{repair work} \cite{jackson2013rethinking} that caseworkers undertake in order to make the algorithm work for their clients. Caseworkers are expected to use algorithmic tools that shift discretion away from them, however, they are also expected to assume responsibility when automated decisions lead to poor outcomes for families. Recently, public sector algorithms have received criticism when they lead to poor outcomes and exacerbate racial disparities \cite{washington2018argue, valentine2019impoverished, eubanks2018automating}. However, finding the sources of harm can be difficult when we witness the distributed use of smaller algorithmic tools as opposed to a larger and more visible algorithmic system. Distributed nature of several of these tools also shifts accountability from one group to another within the agency, and therefore, can be hard to assess. For instance, as highlighted in prior work \cite{saxena2021framework2}, the Anti Sex-Trafficking algorithm refers cases to the Human Anti-Trafficking Response Team (HART) based on a set of risk predictors. However, HART is now receiving an influx of cases with no added resources provided to them. Consequently, cases that really need their expertise are receiving inadequate attention; an indirect harm that is hard to measure. As illustrated in this case study, assessments such as CANS are transformed into algorithmic decision-supports which are now further being legally mandated to ensure caseworkers' compliance. The AFST algorithm implemented at a child welfare agency in Allegheny County, Pennsylvania is another example of this scenario where it was initially implemented in a voluntary capacity, however, its use was mandated when it did not receive enough engagement from caseworkers. AFST is still characterized as an algorithmic decision-support but offers an interesting example of how such tools are increasingly limiting human discretionary work instead of findings ways to improve it.


\section{Implications for Human-AI Decision-Making}
Public sector agencies such as the child welfare agency examined in this study offer an important setting for studying human-AI interaction for the following reasons - 1) full automation in such settings is not desirable because of ethical, legal, and safety concerns associated with high-stakes decision-making \cite{lai2021towards}. This leads to observable ongoing interactions between practitioners and algorithmic systems where practitioners learn to engage with such systems, assess their utility in informing decision-making, as well as develop perspectives regarding critical factors that lead to the development of (dis)trust of such systems, and 2) decisions are collaboratively made by the child welfare staff which allows us to inspect how a team of caseworkers with varying levels of expertise interact and reason with an algorithmic system and incorporate algorithmic decisions within their decision-making processes.

As Lai et al. \cite{lai2023towards} note, the underlying assumption for AI-assisted decision-making is that human reasoning can benefit from counter-intuitive patterns derived from empirical models. Such prediction tasks are also generally referred to as \textit{AI for discovery} tasks \cite{lai2021towards}. For instance, ascertaining the risk of maltreatment as a function of prior involvement with public services (e.g., drug and alcohol services, housing authority, etc.) might help uncover patterns that aid decision-making and prevent future child maltreatment events. However, \textbf{we found that this led to a cumulative distrust of such systems because the decision task was no longer aligned with the caseworkers' practice model and did not help inform their day-to-day decisions.} For instance, caseworkers across jurisdictions \cite{cheng2022child, saxena2021framework2} shared that the knowledge of predicted long-term risk (i.e., the likelihood of being re-referred or placed in foster care within two years) did not help address risk factors that the families experience in the system (e.g., unstable housing or employment, lack of access to medical services, risk posed by procedural factors) and did not help them acquire services for their clients \cite{saxena2023rethinking}. That is, there is a significant difference between the \textit{target variable} that the human decision-makers care about and the \textit{target proxy} being predicted. Guerdan et al. \cite{guerdan2023ground} use causal diagrams to unpack the relationship between predictors, decisions, target variables, and their proxies in the human-AI decision space and show how outcome measurement error occurs due to this gap between the target variable and its proxy. In this study, \textbf{we show how outcome measurement error can result in process-oriented algorithmic harms in practice.}

In addition, for empirical predictions to provide any utility, it is imperative that caseworkers are provided with explanations of predictions that especially highlight these counter-intuitive factors that do not traditionally align with their practice. However, these explanations can differ based on the practitioner and their decision-making goals. Different practitioners had different needs in regard to explainability and a “one-size-fits-all” approach with respect to algorithm design may not be feasible here. For instance, the clinical therapist was more interested in explanations about the context surrounding worsening behaviors whereas the caseworkers wanted explanations about underlying trauma. As Ehsan et al. \cite{ehsan2023charting} note, explainability is a human factor and not just a model-inherent property that requires a careful examination of the sociotechnical gap, i.e. - the divide between the technical affordances and the social needs \cite{ackerman2000}. We direct the readers to Ehsan et al.'s work \cite{ehsan2023charting} where they operationalize a framework for explainable AI (XAI) and highlight how the framework can help researchers better understand the sociotechnical gap in different contexts and practical considerations for how we can begin to bridge this gap.

Most of the \textit{AI for discovery} tasks require domain expertise but domain experts are seldom included in the design of AI systems \cite{lai2021towards}. The majority of studies conducted on human-AI interaction are conducted as isolated experimental trials of decision tasks in lab studies or on crowdsourcing platforms. Consequently, as is evident from the case study of the CANS algorithm, we found that such predicted decision tasks developed without domain expertise did not generalize to decision tasks carried out by caseworkers within the bounds of policies and systemic resources. In addition, seemingly cogent and well-aligned decision tasks can be at odds with one another in practice. In theory, it makes sense to predict the foster care placement setting and the subsidized guardianship rate based on the mental health assessment of foster children. However, in practice, we learned that due to a lack of foster homes in the system, the second-removed outcome (i.e., subsidized guardianship rate) became the primary outcome of interest and impacted how mental health assessments were conducted. \textbf{That is, a systemic issue is the critical driving factor that determines how decision tasks are carried out.} Here, ethnographic engagement is necessary to fully understand the decision-making ecosystem and assess how human-AI decision-making may occur within the bounds of organizational pressures, time constraints, and limited resources and which disruptions are likely to occur in a resource-constrained environment.

In high-stakes domains where improving human-AI interaction is paramount, AI research must be encapsulated within HCI methods because it helps us better understand stakeholders' interactions with the system and assess critical human factors such as trust, reliance, intelligibility, and explainability of the system. In addition, as noted above, AI systems designed for high-stakes domains often seek to replicate or replace human labor without an adequate understanding of the nature of practice and the reality of street-level decision-making. Here, HCI methods can help us understand organizational workflows, day-to-day protocols that workers are expected to follow, systemic factors (e.g., policies, resources, etc.), and the bottlenecks in decision-making. Consequently, this helps us outline the systemic mechanics as well as the insertion points, scope, and utility of algorithmic decision-making. That is, designing systems that algorithmically support the practitioners' cognitive environment with or without predicting decisions. The 7ei algorithm provides an interesting case study of a simple algorithmic tool specifically designed to overcome the obstacles described above. Having faced several frustrations with empirical models, the agency leadership engaged in theory-driven design and developed 7ei using the principles of trauma-informed care (TIC). This further allowed them to ensure that decisions made using 7ei are centered in the theory of practice. \textbf{The agency also ensured that child welfare staff were trained in TIC, created a collaborative setting to ensure the proper use of the tool, and decided to track multiple TIC outcomes over the life of cases instead of predicting them}. Consequently, what we observed over an extended ethnography was that 7ei received collective buy-in and significantly better engagement from the caseworkers because the tool was centered in social work practice. 

As highlighted by Vaughan and Wallach \cite{vaughan2020human}, it is imperative to understand the needs of relevant stakeholders to be able to design for intelligibility techniques that meet these needs. It is imperative to note here that 7ei was designed to cultivate the human factors of trust, reliance, and intelligibility. It facilitates the child welfare staff's intelligibility needs by 1) providing relevant information, 2) centering this information in a theoretical framework and 3) demonstrating compliance with evidence-based practice. The tool was also designed with the intent to decompose the algorithm and turn it into an open-ended and transparent process that promotes trust and reliance. The collaborative use of 7ei also helped establish informed mistrust of the tool for cases where decisions were dictated by policy and/or systemic constraints with less utility for TIC principles. However, it is necessary to note that too many interactions where the tool did not inform decision-making (due to persisting systemic factors) may lead caseworkers to distrust and abandon the tool altogether. Here, from a human-AI interaction standpoint, it is imperative to note that \textbf{designing for intelligibility needs preceded designing for explainability or transparency and impacted the caseworkers' engagement and reliance on the tool}. Transparency and explainability are often endorsed as human factors that might facilitate engagement and improve decision-making (i.e., making the workings of the algorithm transparent and providing explanations for predictions). Here, the case study of the CANS algorithm provides a cautionary tale against prioritizing these human factors over the intelligibility needs of stakeholders. Even though CANS promotes transparency where all the predictors are accessible to caseworkers, it frustrates the child welfare staff because it does not account for the principles of TIC and only offers an isolated view of the child. This further problematizes explainability because caseworkers were more invested in explanations of underlying trauma than exhibited behaviors. Consequently, we learned that transparency in the case of CANS promoted the manipulation of data, led to the diminishing of trust and reliance, and forced caseworkers to engage in repair work to make the tool work for their clients.


\section{Implications for Responsible Data Science in the Public Sector}

\begin{figure}
  \includegraphics[scale=0.45]{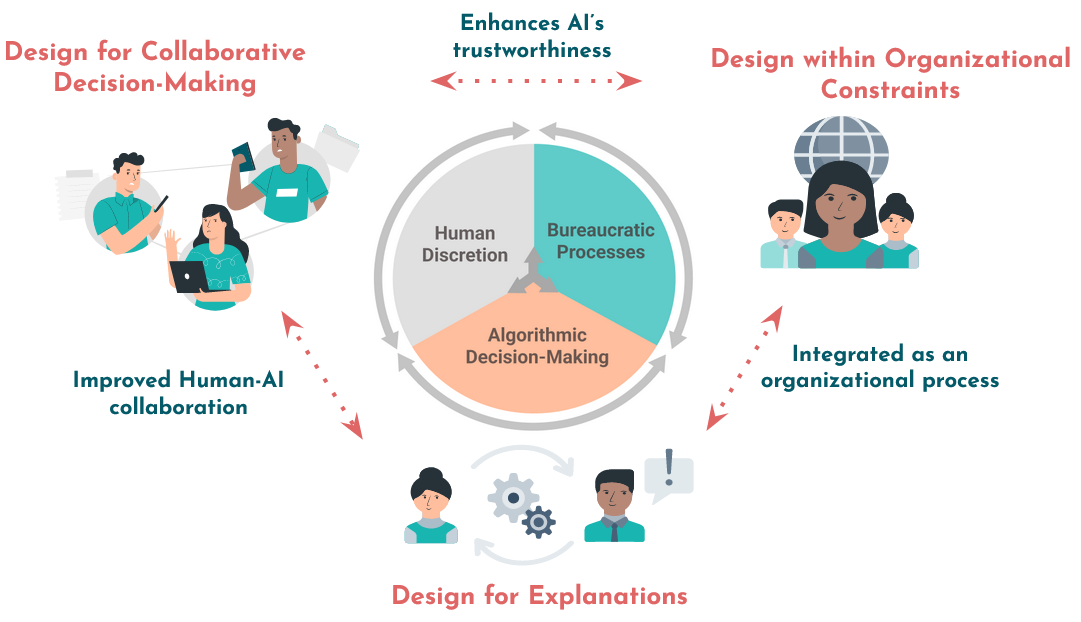}
  \caption{Designing Algorithms for the Public Sector through the Lens of ADMAPS Framework}
  \vspace{-0.3cm}
  \label{fig:7eioutcomes}
\end{figure}

Crucial to the discussion of responsible data science practices \cite{aragon2022human} in the public sector are the following points. First, it is imperative to recognize that decision-making is a complex process where information needs to be shared among several parties (e.g., child welfare staff, district attorney's office, parents' attorneys, service providers, and judges) and decisions are collaboratively made. As illustrated by Figure 2, \textbf{algorithmic tools must be designed for collaborative use} for them to offer higher utility to practitioners and improve their decision-making processes. In addition, practitioners must be able to explain decisions made to other involved parties. Therefore, it is necessary to \textbf{design algorithms that facilitate explanations}. Here, Ehsan et al's framework for operationalizing explainable AI \cite{ehsan2023charting} offers important practical considerations for designing for explanations when the nature of practice is highly collaborative. Designing for collaborative use and explanations will aid deeper integration between human discretion and algorithmic decision-making, and consequently, lead to improved human-AI decision-making over time. In addition, \textbf{algorithm design must occur within organizational constraints}, i.e.- it is imperative to account for the resource and legislative constraints within which all decisions (human or algorithmic) must be made. In the absence of these design requirements, algorithms frustrate practitioners who are unable to follow the `ideal algorithmic decision'. For instance, CANS might recommend that a foster child be placed in a Level 4 foster home, however, with a lack of good foster homes in the system, the caseworker must manipulate data to produce an outcome that points to an available placement \cite{saxena2020child}. Consequently, algorithmic tools that account for such constraints are more likely to be integrated as an organizational process as well as enhance the tool's trustworthiness among practitioners who must interact with it collaboratively.

Second, as highlighted by the adoption and use of the 7ei algorithm, agency leaders had to invest a significant amount of resources in terms of trainings, specialized consultations, hiring experts, and creating the time and space (in terms of collaborative 7ei meetings) to ensure proper utilization of the algorithmic tool. That is, there is a significant amount of human labor and agency resources that went into the integration of 7ei into daily work routines and decision-making processes. It is imperative to note that these investments, especially in the public sector, must be made in order to deconstruct and rebuild decision-making processes that utilize algorithmic systems. 

Third, as highlighted by the development of 7ei, the tool was designed with the intent to decompose the algorithm and turn it into an open-ended and transparent process such that it tracked outcomes over time instead of predicting an outcome of interest. As highlighted by prior work, there is an irreducible degree of uncertainty associated with each predicted outcome \cite{de2020case, paakkonen2020bureaucracy} and this problem is further exacerbated in the public sector where all the relevant information is not always available and there may also be contradicting sources of information \cite{bullock2019artificial, saxena2022unpacking}. In such scenarios, predicted decisions led to frustrations on the part of caseworkers because there is a significant gap between the \textit{unobserved target variable} and the \textit{observed proxy variable} of interest to the human decision-makers \cite{guerdan2023ground}. In sum, there is a need to prescribe away from the black-box model of \textit{\textbf{input-computation-output}} and build tools that algorithmically support the practitioners' cognitive environment without predicting decisions \cite{saxena2023rethinking, saxena2022unpacking}. Designing to support decision-making processes will also lead to fair, transparent, and accountable decisions for clients as well as responsible AI practices from the data science community. Furthermore, as highlighted by prior work \cite{vaughan2020human}, there is no need to over-engineer solutions using complex neural networks when simple models have been shown to be just as accurate in high-stakes domains \cite{vaughan2020human, jung2017simple, rudin2019stop}. In addition, caseworkers' frustrations with CANS and the repair work they conducted highlighted key aspects of decision-making that the algorithm failed to capture. This further highlights the need to conduct ethnographic work as well as the need to co-design with stakeholders such that their needs and concerns are addressed. Here, we direct the readers to Yildirim et al.'s work \cite{yildirim2023creating} on how to effectively ideate AI concepts and explore AI's problem-solution space with domain experts.  


\section{Limitations}
We conducted an extensive ethnographic study that highlights the messy interactions between social work practice, policies, and algorithmic decision-making at a child welfare agency. However, this study has some limitations that create opportunities for researchers to further expand upon this body of work. First, this study only focuses on the perspectives of on-the-ground caseworkers and on their interactions with algorithmic systems. It is important to understand the perspective of affected communities (i.e., foster children, parents, and foster parents) about whom decisions are being made through algorithms. For instance, a recent study conducted by Stapleton et al. \cite{stapleton2022imagining} found that parents considered CWS algorithms to be punitive and unsupportive. Parents instead wanted systems that would help them fight against CPS as well as evaluate CPS and the caseworkers themselves. Future research should continue to focus on uncovering street-level complexities within this complicated sociotechnical environment to understand how different labor practices, systemic constraints, and power asymmetries might amplify algorithmic harms experienced by citizens on the street level \cite{saxena2022unpacking, saxena2023rethinking}. Finally, this ethnographic study only uncovers complexities at one child welfare agency, however, child welfare practices and policies can significantly vary from one state to another. Therefore, we recommend that researchers conduct similar qualitative explorations in other jurisdictions.


\section{Conclusion}
We conducted an in-depth ethnographic study to understand the daily algorithmic practices of caseworkers at a child welfare agency. We qualitatively coded our data from the ethnography to the dimensions of the ADMAPS framework to reveal the complex interdependencies between human discretion, algorithmic decision-making, and bureaucratic processes. We focused on the macro-interactions between the three core elements of ADMAPS to understand how the interplay between systemic mechanics and algorithmic decision-making impacts the fairness of the decision-making process itself. Our findings highlight how functionality issues in algorithmic systems can lead to process-oriented harms where they adversely impact the nature of professional practice, and administration at the agency, and lead to inconsistent and unreliable street-level decision-making. Here, caseworkers are compelled to undertake additional labor in the form of repair work to restore disrupted administrative processes and decision-making at the street level, all while facing organizational pressures and time and resource constraints. We learned that there is a need to focus on the proper implementation and integration of algorithmic tools into decision-making processes and not just the initial development and deployment of the algorithmic model. That is, there is a need to rethink the amount of investment required to ensure the proper adoption of algorithms in complex sociotechnical environments. In addition, we show how a simple algorithmic tool that tracks variables over time instead of predicting an outcome offered higher utility to caseworkers. Here, algorithms need to be designed to support explanations and collaborative use such that they augment human discretionary work without seeking to replace or replicate it. In addition, algorithmic tools need to be fully supported by bureaucratic processes by allocating necessary resources and accounting for organizational constraints to ensure that they are integrated as an organizational process. As a result of this study, we also offer implications for responsible data science practices in the public sector.


\begin{acks}
This work was initially presented at Data \& Society's "The Social Life of Algorithmic Harms" academic workshop held in March 2022. We sincerely thank the organizing team at Data \& Society as well as the participants and reviewers for their valuable suggestions, detailed feedback, and most of all for their intellectual generosity.    
\end{acks}


\bibliographystyle{ACM-Reference-Format}
\bibliography{bibliography}
\end{document}